\documentclass[aps,pre,twocolumn,superscriptaddress,showpacs,amsmath,amssymb,floatfix]{revtex4-1}

\usepackage{graphicx}
\usepackage{color}
\usepackage{verbatim}
\usepackage{epsfig}
\usepackage{dcolumn}
\usepackage{bm}
\usepackage{amsmath}%
\usepackage{amssymb} 
\usepackage{enumitem}

\newcommand{\invt}{\ensuremath{\tau_0^{-1}}}

\begin{document}
\title{A computational study of transient shear banding in soft jammed solids}
\author{Vishwas V. Vasisht}
\author{Emanuela Del Gado}
\affiliation{Department  of  Physics,  Institute  for Soft  Matter  Synthesis and Metrology,
Georgetown  University,  37th and O Streets,  N.W., Washington,  D.C. 20057,  USA}
\date{\today}
\begin{abstract}
We have designed 3D numerical simulations of a soft spheres model, with size polidispersity and in athermal conditions, to study the transient shear banding that occurs during yielding of jammed soft solids. We analyze the effects of different types of drag coefficients used in the simulations and compare the results obtained using Lees-Edwards periodic boundary conditions with the case in which the same model solid is confined between two walls. The specific damping mechanism and the different boundary conditions indeed modify the load curves and the velocity profiles in the transient regime. Nevertheless, we find that the presence of a stress-overshoot and of a related transient banding phenomenon for large enough samples are a robust feature for overdamped systems, where their presence do not depend on the specific drag used and on the different boundary conditions. 
\end{abstract}

\maketitle

\section{Introduction}

Solids whose microscopic constituents are densely packed into an amorphous assembly and basically insensitive to thermal fluctuations form an integral part of our everyday life, with examples ranging from pharmaceutical, cosmetic and food products to 
wet cement. Hence understanding the flow properties of such materials is of technological importance \cite{bibette_Springer_2003, vanHecke_JPCM_2009, Bonn_RMP_2017}. Under shear deformations, these materials flow only when the applied shear stress is above a threshold value, the yield stress. As the imposed strain increases at a fixed shear rates, they show an initial linear increase in the stress, often followed by a stress overshoot, beyond which the system yields before reaching a steady flow state \cite{varnik_JCP_2004, fielding_RPP_2014, Divoux_ARFM_2016, shrivastav_JOR_2016}. The steady state shear stress ($\sigma$) as a function of the applied shear rate ($\dot{\gamma}$) provides a constitutive behavior (flow curve) that in most cases is  well described by the Herschel-Bulkley (HB) curve $\sigma = \sigma_Y + \kappa \dot{\gamma}^n$, where the yield stress $\sigma_Y$, the effective viscosity $\kappa$ and the exponent $n$ are in principle material specific \cite{hebraud_prl_1998, nicolas_SM_2014}. While the constitutive behavior under flow is interesting in itself and is the subject of many investigations \cite{Goyon_Nat_2008, lemaitre_prl_2009, Tighe_PRL_2010, martens_SM_2012, Seth_NMat_2011, vasisht_prl_2018, Nicolas_RMP_2018}, here we are interested instead in the transient behavior of the material before it reaches the steady-state flow, where it maintains features of both its solid and fluid response. Before reaching the steady flowing state, jammed soft solids often display flow instabilities which manifest in terms of strong spatial inhomogeneities in the flow profile, often called shear localization or shear banding, even when the material is homogeneously driven. 

Shear banding in complex fluids like polymer solutions and wormlike micelles has been extensively studied and, in many cases, its origin is understood in terms of flow alignment of the microstructure or flow induced crystallization, through theoretical approaches that couple the flow fields with the microstructure \cite{dhont1999constitutive, olmsted_rheoacta_2008, dhont2008gradient, lerouge2009shear, Bonn_RMP_2017}. In jammed and soft glassy materials, these phenomena are the subject of intense investigations \cite{becu_PRL_2006, fielding2009shear, divoux_prl_2010, besseling_PRL_2010, mansard2011kinetic, martens_SM_2012, fielding_RPP_2014, irani_PRL_2014, Divoux_ARFM_2016, Bonn_RMP_2017, gross2018shear} and their possible association to an underlying non-equilibrium phase transition is increasingly debated \cite{moorcroft_PRL_2011, adams2011transient, wisitsorasak2017dynamical, parisi2017shear, ozawa2018random, popovic2018elastoplastic}. However a good understanding of their microscopic origin is fundamentally lacking and
particularly elusive when the shear banding is observed as a transient feature that does not persist in steady-state \cite{fielding_RPP_2014, divoux_prl_2010, Divoux_ARFM_2016, shrivastav_JOR_2016}.

Microscopic computer simulations can be effective in bringing new insight, but most of the existing studies have been performed in 2D and in the quasi-static limit (i.e. zero shear rate) while experiments are always performed at finite shear rates and nearly always in 3D.  Together with meso-scale simulations based on elasto-plastic models \cite{martens_SM_2012, Nicolas_RMP_2018}, most of existing microscopic simulation studies have focused on the part of the phenomenon that can be rationalized in terms of the emergence of local plastic rearrangements in an elastic background \cite{lemaitre_prl_2009}. That is, they do not capture the coupling of the microscopic dynamics to the imposed deformation rate. 
Recent works, instead, have pointed precisely to the crucial role of the rate dependent dynamics, calling for dedicated numerical investigations \cite{divoux_prl_2010, Divoux_SM_2011, fielding_RPP_2014}.

To be able to address such questions, we have devised a numerical study using 3D microscopic simulations and finite deformation rates. In this paper we analyze the outcomes for the shear start-up response under varying shearing protocols and boundary conditions, with the goal to extract the robust features that do not depend on the specifics of the numerical simulations. We study a jammed suspension (at  volume fraction $\phi\simeq0.7$) of polydisperse spheres with soft repulsive interactions, i.e., a deeply jammed soft amorphous solid. We find that the shear start-up is characterized by a stress-overshoot and formation of a transient band, which eventually disappears upon reaching the steady state. We extensively test whether such findings depend on different implementations of the viscous drag due to the background solvent or on different types of boundary conditions. Our study indicates that the stress-overshoot and the shear banding during the shear start-up are robust features emerging from the microscopic physics of our model material, they do not qualitatively change with the different shearing protocols used here and are mainly controlled by the sample age that we can change through the preparation protocol. 

The paper is organized as follows. In section II we discuss the model system and the sample preparation protocol as well as the rheological and structural characterization of the initial sample configurations. In section III we discuss the shearing protocols followed in our work for performing finite shear rate simulations. Sections IV and V contain the results comparing the transient rheological properties obtained from different protocols. At the end, in section VI we present conclusions and discussion.

\section{Model and sample preparation}
The 3D model for soft amorphous material we investigate is composed of a non-Brownian suspension of soft repulsive particles interacting via a truncated and shifted Lennard-Jones potential \cite{WCA_JCP_1971}, given by $U(r) = 4\epsilon \left [(a_{ij}/r_{ij})^{12} - (a_{ij}/r_{ij})^{6} \right ] + \epsilon$, for $r_{ij} \le 2^{1/6} a_{ij}$, else $U(r_{ij})=0$. Here $\epsilon$ is the unit energy in the simulations, $r_{ij}$ being the center to center distance between the particle $i$ and $j$ and $a_{ij}=0.5(a_i+a_j)$, with $a_i$ and $a_j$ being the diameter of particles $i$ and $j$ respectively. The diameters of the particles are drawn from a Gaussian distribution with variance of 10\%, whose mean is used as unit length $a$. All the simulations are performed at a volume fraction $\phi \approx 70\%$, consisting of $10^{5}$ (97556) particles (unless otherwise mentioned). Albeit simple, this type of model has been successfully used for numerical simulations of soft solids in various contexts, and proven to capture the fundamental physics of their behavior under deformation \cite{varnik_PRL_2003, irani_PRL_2014, shrivastav_JOR_2016, vasisht_prl_2018}.

We prepare the numerical samples to be sheared using the following procedure. An initial FCC crystal at the chosen volume fraction of $0.70$ (with $lx=ly=lz=42.1798 a$) is melted at $T=5.0 \epsilon / k_B$ and equilibrated at the same temperature using NVT Molecular Dynamics (MD) for  $\simeq 5\cdot 10^{4}$ MD steps, with a MD timestep of $\Delta t = 0.001 \tau_{0}$ (where $\tau_{0}=\sqrt{ma^{2}/\epsilon}$ is the unit of time and $m$ is the unit mass). After making sure that there is no signature of crystallinity in the equilibrated sample by measuring the bond orientational order parameter $Q_6$ \cite{steinhardt_PRB_1983}, the melt is subjected to a systematic temperature quench. From the initial temperature of $T=5.0 \epsilon / k_B$ we decrease the system temperature by $\Delta T$, after which we let the system relax at the temperature $T -\Delta T$ for $10^{4}$ MD steps. We repeat this procedure several times, until $T=0.001 \epsilon/k_B$ is reached. By changing the $\Delta T$ we control the cooling rate $\Gamma$. The samples discussed here have been prepared for $\Gamma$ corresponding to $5 \cdot 10^{-2}, 5 \cdot 10^{-3}, 5 \cdot 10^{-4} \epsilon/(k_B \tau_0)$. After the system reaches $T=0.001 \epsilon/k_B$, we perform an energy minimization using the conjugate gradient (CG) method to take the system to the zero temperature limit. For the fastest quench rate case, we minimise the energy directly from the equilibrium liquid state at $T=5.0 \epsilon/k_B$ using CG. Following the above described protocol we prepare five independent samples at each cooling rate (except for the lowest one, where we have three samples). The initial configurations so obtained are subjected to shear deformation at finite shear rate. Before studying the shear start-up behaviour we carefully analyze various properties of the initial sample configurations.

\begin{figure}[h]
\includegraphics[width=0.400\textwidth]{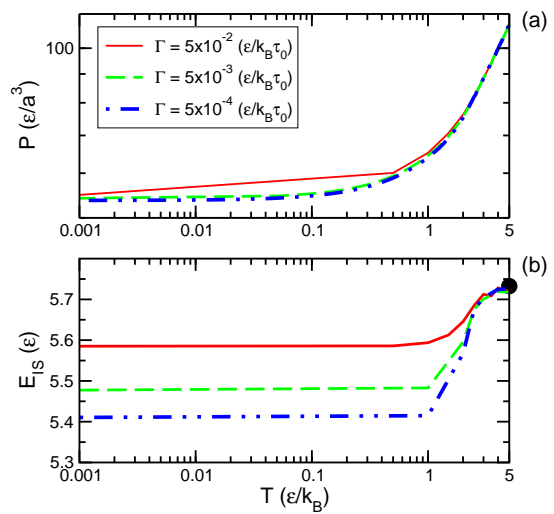}
\caption{(a) Pressure computed as a function of applied temperature for samples quenched at different cooling rates. The inset shows the zoomed-in part near the final applied temperature ($T=0.001 \epsilon/k_B$). (b) The inherent structure energy or the energy associated with the nearest local minimum in the potential energy landscape, as a function of temperature for different cooling rates. With the decrease in the cooling rate, one explores the deeper basins of the potential energy landscape.}
\label{Fig1}
\end{figure}

In Fig. \ref{Fig1}(a) we show the pressure measured as a function of the temperature as the sample is quenched at a chosen cooling rate $\Gamma$. The pressure dependence on the temperature does not change with $\Gamma$ until $T=3.0 \epsilon/k_B$, which in our model corresponds to the onset of slow glassy dynamics. Below such onset temperature, the pressure achieved through our cooling protocol decreases with decreasing $\Gamma$. In Fig. \ref{Fig1}(b), we plot the energy of the closest local minimum accessible to the system, obtained through CG, as a function of $T$. One can clearly see that, upon decreasing $\Gamma$, we access deeper local minima (or inherent structures) of the total potential energy \cite{sastry_nat_1998,  stillinger1995topographic, mosayebi2014soft, ashwin2004relationship} below the onset temperature. We next characterize the mechanical and structural properties of these inherent structure configurations obtained from difference $\Gamma$.


\subsection{Linear viscoelastic response of the initial configurations}
In order to characterize the mechanical properties of the initial configurations, we compute the complex modulus by performing small amplitude oscillatory shear. By applying a shear strain $\gamma(t) = \gamma_0 sin(\omega t)$, for a strain amplitude of $\gamma_0 = 1\%$, we monitor the stress response for varying frequencies $\omega$ \cite{larson1999structure}. We monitor the energy and pressure evolution during the oscillatory shear cycles and once the system reaches a saturation in these quantities as a function of the number of cycles, we extract the linear visco-elastic moduli (respectively storage and loss) using 
\begin{eqnarray}
G'(\omega, \gamma_0) = \frac{\omega}{\gamma_0 \pi} \int_{t_0}^{t_0 + 2\pi/\omega} \sigma_{xy}(t) sin(\omega t) dt,\\
G''(\omega, \gamma_0) = \frac{\omega}{\gamma_0 \pi} \int_{t_0}^{t_0 + 2\pi/\omega} \sigma_{xy}(t) cos(\omega t) dt
\end{eqnarray}
\noindent 

\begin{figure}
\includegraphics[width=0.400\textwidth]{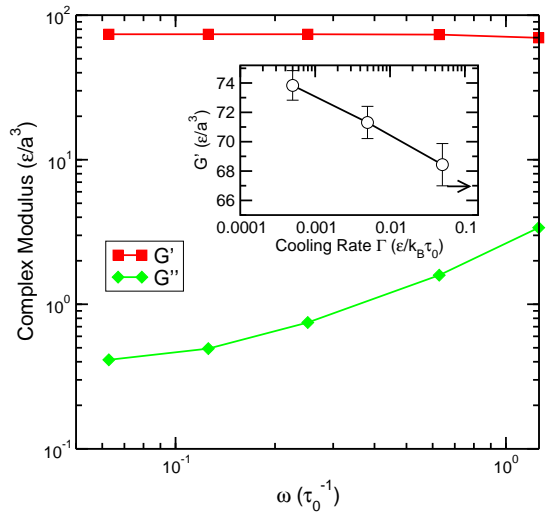}
\caption{(main panel) The complex moduli (elastic - G' and plastic - G'') computed as a function of frequency for an intial configuration prepared at a cooling rate $\Gamma=5 \cdot 10^{-4}$. (inset) The zero frequency storage modulus G' as a function of cooling rate $\Gamma$. The G' increases with decreasing the cooling rate. The arrow indicates the value computed for the infinitely fast cooling rate, obtained from the CG minimization of a high T configuration.}
\label{Fig2}
\end{figure}

In Fig. \ref{Fig2} we show the $G'(\omega, \gamma_0=0.01)$ and $G''(\omega, \gamma_0=0.01)$ as a function of the applied frequency $\omega$ for the initial configuration prepared at a cooling rate $\Gamma=5 \cdot 10^{-4} \epsilon/k_{B}\tau_{0}$. The data show that the sample is solid, with $G' >> G"$ at low frequencies and the zero frequency storage modulus can be extracted by extrapolating the data to $\omega=0$.  We compute the zero frequency storage modulus for samples prepared with different $\Gamma$, as shown in the inset of Fig. \ref{Fig2}. The arrow in the inset of Fig. \ref{Fig2} corresponds to the fastest cooling rate, where a liquid configuration is directly subjected to CG minimization. With decreasing the cooling rate $\Gamma$, the configurations in deeper local minima of the potential energy landscape, as identified in Fig. \ref{Fig1} (b), clearly correspond to solids with higher mechanical strength.

\subsection{Structural analysis}
\begin{figure}
\includegraphics[width=0.400\textwidth]{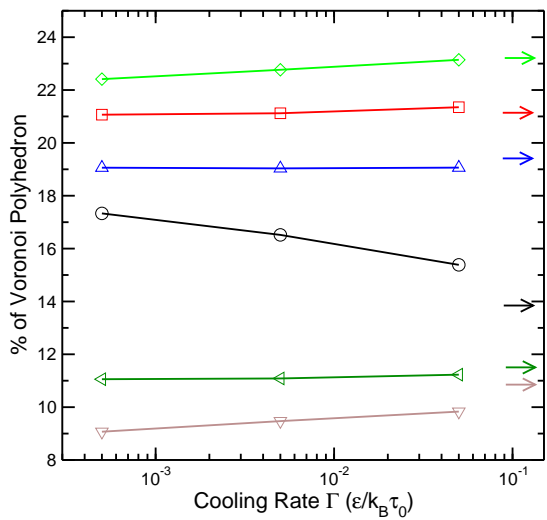}
\caption{Fraction of different Voronoi polyhedra in a sample prepared using different cooling rates. In each initial configuration, a Voronoi polyhedron is identified from the number of faces, vertices and edges, using a Voronoi tesselation analysis. Any polyhedron species that corresponds to at least  $5\%$ of the total is considered. Open circles correspond to Face(F)=12, Edges(E)=30 and Vertices(V)=20, a Voronoi dodechedron or a particle with Icosahedron neighbors. Likewise, squares (F=13;E=33;V=22), diamonds (F=14;E=36;V=24), upside triangles (F=15;E=39;V=26), left-side triangles (F=16;E=42;V=28) and down-side triangles (F=17;E=;V=30). The star denotes the data for the infinitely fast cooling rate, obtained from conjugate gradient minimization. Compared to other polyhedra, clearly the icosahedra fraction increases with decreasing cooling rate.}
\label{Fig3}
\end{figure}

In order to gain insight into the microscopic structural underpinnings, we analyze the structure of the samples by constructing a 3D radical Voronoi tesselation 
with the {\it Voro++} open source software library \cite{rycroft_chaos_2009} and compute the Voronoi polyhedra. This method is well suited for our samples made of polydisperse spheres. For samples prepared at different cooling rate $\Gamma$, we obtain the statistics of various Voronoi polyhedra (defined by number of faces, edges and vertices). In Fig. \ref{Fig3}, we show the fraction of different polyhedra (which constitute more than $5\%$ of each sample) as a function of the cooling rate. The data show that the fraction of Voronoi dodecahedra is clearly affected by the cooling rate $\Gamma$, whereas the contribution of the other polyhedra does not change much. A Voronoi dodecahedron corresponds to a local packing which is an icosahedron, where neighbouring particles form an icosahedron around a reference particle. Relatively higher percentages of local icosahedral packing are expected and widely observed in slowly quenched supercooled liquids \cite{steinhardt_PRL_1981}, as well as in glasses, with similar spherically symmetric interactions \cite{mosayebi2012deformation, royall_PR_2015, ronceray2015favoured}.

Having characterized the rheological and structural properties of the initial configurations prepared at the various cooling rates, we now analyze the rheological response of a sample obtained with the slowest cooling rate, by subjecting it to a continuous shear deformation at a finite rate. 

\section{Shearing protocols}

The samples characterized as described above are subjected to a shear deformation of strain amplitude $\gamma (t)$ with constant imposed rate $\dot{\gamma}$, either using Lees-Edwards boundary conditions (LEBC) or using a wall-based protocol (WB). Note that in all simulations discussed below [x,y,z] refers respectively to flow, gradient and vorticity directions. In the WB protocol, we confine the sample in $y$ direction by freezing two layers of particles at the simulation box boundary. $L_{y}$ is the width of the box in the $y$ direction for the confined samples. All shear rates are measured in units of $\invt$.

For LEBC we solve the equations of motion in two different ways.  
In the first way, which in the following we call LEBC1, we use the following dissipative particles dynamics (DPD) equations of motion \cite{nicolas_PRL_2016}: 
\begin{equation}
m \frac{d^2\vec{r}_i}{dt^2} = - \zeta_{DPD} \sum_{j(\ne i)} \omega(r_{ij}) (\hat{r}_{ij}\cdot \vec{v}_{ij})\hat{r}_{ij} - \triangledown_{\vec{r}_i} U
\label{DPD}
\end{equation}
\noindent where $m$ is the mass of the particle and the first term in the right hand side (RHS) is the damping force which depends on the damping coefficient $\zeta_{DPD}$. The relative velocity $\vec{v}_{ij} = \vec{v}_j - \vec{v}_i$ is computed over a cut-off distance $r_{ij} \le 2.5 a_{ij}$, with the weight factor $\omega(r_{ij})=1$. These choices for the cut-off distance and the weight factor are consistent with other studies in the literature for similar systems \cite{Xu_PRL_2005, maloney_JPCM_2008, Puosi_SM_2015, nicolas_PRL_2016}. In the viscous damping we have only considered the radial contribution of the relative velocities, since the particles are point-like and the main sources of change in the velocity are the inter-particle forces which are purely radial \cite{puosi_PRE_2014}. Nevertheless, the tangential contribution to the damping forces can also be included as $((\hat{r}_{ij}\cdot \vec{v}_{ij})\hat{r}_{ij}) + (\vec{v}_{ij} - (\hat{r}_{ij}\cdot \vec{v}_{ij})\hat{r}_{ij})$ \cite{baumgarten_SoftMatter_2017}. In most of this work we have used the formulation with only the radial contribution, but in one of the following sections we briefly discuss the effect of including the transverse term, which has also been investigated recently in \cite{Irani_PRF_2019}. The second term in the RHS is the force due to the interactions between particles.

In the second way, which we indicate as LEBC2, we solve equations of motion where the solvent drag is Stokes-like and only depends on the absolute particle velocity \cite{ikeda2013disentangling, colombo2014stress, Hoh_JFM_2015, Su_JCP_2017}, as in a free draining approximation, given by \begin{equation}
	m \frac{d^2\vec{r}_i}{dt^2} = -\zeta_{SD} \left (\frac{d\vec{r}_i}{dt} - \dot{\gamma} y_i \vec{e}_x \right ) - \triangledown_{\vec{r}_i} U
\label{SD}
\end{equation}
\noindent where again $m$ is the particle mass and the first term in the RHS is the damping force which depends on the damping coefficient $\zeta_{SD}$. Hence the drag force is proportional to the difference between the particle's velocity $d\vec{r}_i/dt$ and an affine background velocity, dictated by imposed shear rate, given by $\dot{\gamma}y_i\vec{e}_x$. The second term in the RHS is the force due to the interactions between particles.
While one concern with this second approach is that it does not strictly conserve momentum (i.e., it is not Galileian-invariant) \cite{maloney_JPCM_2008, Soddemann_PRE_2003}, its use can still be justified in systems where most of the stress induced through the imposed deformation is due to inter-particle interactions and the contribution of the solvent is a minor correction, since this type of drag term allows for simpler and faster simulations, with no need to adjust additional parameters such as the cut-off or the form of the weight factor in the DPD one. We will show in the following that this free draining approximation works well in the case of deeply jammed systems as those of interest here.  

\begin{figure}[t]
\includegraphics[width=0.500\textwidth]{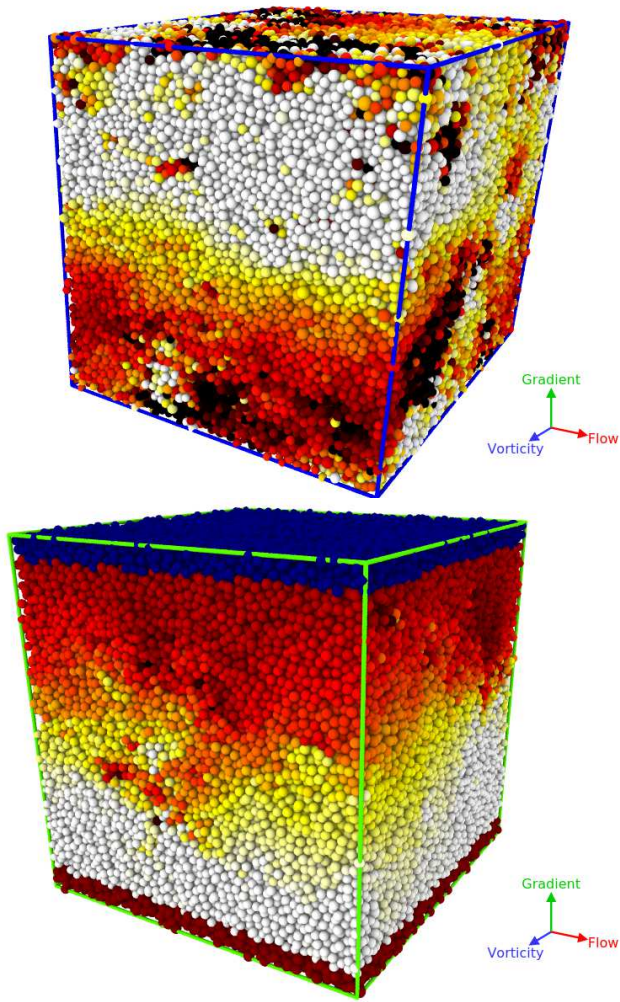}
\caption{Simulation rendering of  samples prepared at $\Gamma=5\cdot10^{-4}\epsilon/(k_{B} \tau_{0}$ and sheared at $\dot{\gamma} 10^{-4} \invt$ using LEBC (top) and wall based (bottom) protocol. In both cases, we have used the DPD based protocol with $Q=1$. We use the hot colour gradient scheme based on the velocities of the particles which vary between 0 $a/\tau_{0}$ (white) and 0.005 $a/\tau_{0}$ (black). In the wall based protocol the top wall (dark blue particles) is moved at chosen velocity associted with the applied shear rate, while the bottom wall (maroon particles) is kept fixed.}
\label{Fig4}
\end{figure}

For the wall based shear deformation tests, which we call WB in the following, we follow the procedure of Varnik and co-workers in \cite{varnik_JCP_2004}. Two walls confine the samples along the direction $\hat{y}$, at a relative distance $L_{y}$: one wall moves at a velocity $\vec{v} = v_{x}^{wall} \hat{x} = \dot{\gamma}L_y \hat{x}$), while the other is kept fixed. The particles that form the wall are completely frozen during the evolution of the system, but the interactions between the wall and the sample particles are the same as in the sample. For the WB shear deformation we use only the DPD approach as in eqn.(\ref{DPD}). In fact, if the drag coefficient is directly proportional to the particle velocity as in eqn.(\ref{SD}), in the WB simulations the particles near the moving wall feel more drag than the ones away from it. As a consequence, for large samples the time needed for particles near to the non-moving wall to sense the deformation is quite a long (i.e., quite larger than the sound speed in the system). This behavior could be mistaken for a shear banding but it is only a numerical artifact due to the use of eqn.(\ref{SD}) in this geometry and disappears using instead the DPD equation, eqn.(\ref{DPD}). 

Finally, both eqs. \ref{DPD} and \ref{SD} include the particle inertia since this allows us to use the same efficient and precise algorithms devised for MD \cite{FrenkelandSmit}. Nevertheless we can effectively study the dynamics in the overdamping limit regime by suitably adjusting the damping coefficients $\zeta_{SD}$ and $\zeta_{DPD}$. As a measure of the extent of damping, we can define an inertial quality factor $Q=\tau_{\mathrm damp}/\tau_{\mathrm vib}$, that measures the ratio between the time scale over which the inertial motion is damped $\tau_{\mathrm damp}=m/\zeta$, due to the solvent drag, and the time scale $\tau_{vib}=\sqrt{m a^2 /\epsilon}$ over which a particle of mass $m$ and diameter $a$ is accelerated by the force $\epsilon/a$. For a fully over-damped system $Q\rightarrow0$. It has been shown that for the athermal conditions considered here, and for similar type of interactions, $Q\approx1$ well approximates the overdamped regime \cite{nicolas_PRL_2016, vasisht2018permanent}. Hence here we focus on $Q\approx1$ and show also that the results we obtained do not vary significantly by decreasing $Q$ or varying it around $1$. As stated above, the shear rates in the following are always expressed in units $\invt \equiv \tau_\mathrm{vib}^{-1}$.




All simulations have been performed using LAMMPS molecular dynamics package \cite{LAMMPS_JCP_1995}, with modifications to the source code to incorporate continuous polydisperse interactions, since we use a distribution of particle sizes and the interaction shape depends on the particle diameters. In the Fig. \ref{Fig4}, we show a rendering of samples prepared at $\Gamma=5 \cdot 10^{-4} \epsilon/(k_{B} \tau_{0})$ and sheared at $\dot{\gamma} = 10^{-4} \invt$ (the snapshots are both taken at $\gamma=0.12$, during the stress decay) using LEBC and wall based protocols. We use the color scheme to show the velocities of the particles which vary between 0 $a/\tau_{0}$ (white) and 0.005 $a/\tau_{0}$ (black).

In the following section we present a comparative study of the rheological response starting from the load curves, using the two different boundary conditions and the two different equations of motion introduced above. 

\section{Load curves and flow profiles: comparison of the different shearing protocols.}

We begin with the rheological data obtained for a sample prepared at cooling rate $\Gamma=5\cdot10^{-4} \epsilon/(k_{B} \tau_{0})$, i.e., a well annealed sample, and for a sample obtained by directly minimizing the energy of the high temperature fluid with CG, i.e., a poorly annealed sample. Both samples are sheared at $\dot{\gamma}=10^{-4}\invt$ and $\dot{\gamma}=10^{-2}\invt$, using the different protocols discussed above. All data in this section corresponds to the  overdamping limit, with the inertial quality factor $Q =1$. Since all the simulations are performed in athermal conditions, the shear stress is computed from the virial stress tensor $\sigma_{\alpha \beta} = \frac{1}{V} \sum_{i} \sum_{j>i} r_{ij}^{\alpha} f_{ij}^{\beta}$, where V($=lx*ly*lz$) is the volume of the system, $r_{ij}$ represents the distance between particle $i$ and $j$, $f_{ij}$ is the force on the particle $i$ due to particle $j$ and ${\alpha,\beta} \in {X,Y,Z}$. We indicate the shear component of the stress $\sigma_{XY}$ with $\sigma$ and the virial pressure is obtained as $P=\frac{1}{3} \left(\sigma_{XX}+\sigma_{YY}+\sigma_{ZZ} \right )$. The first  normal stress difference is computed as $\sigma_{11} = \sigma_{XX} - \sigma_{YY}$.
\begin{figure*}
\includegraphics[width=0.900\textwidth]{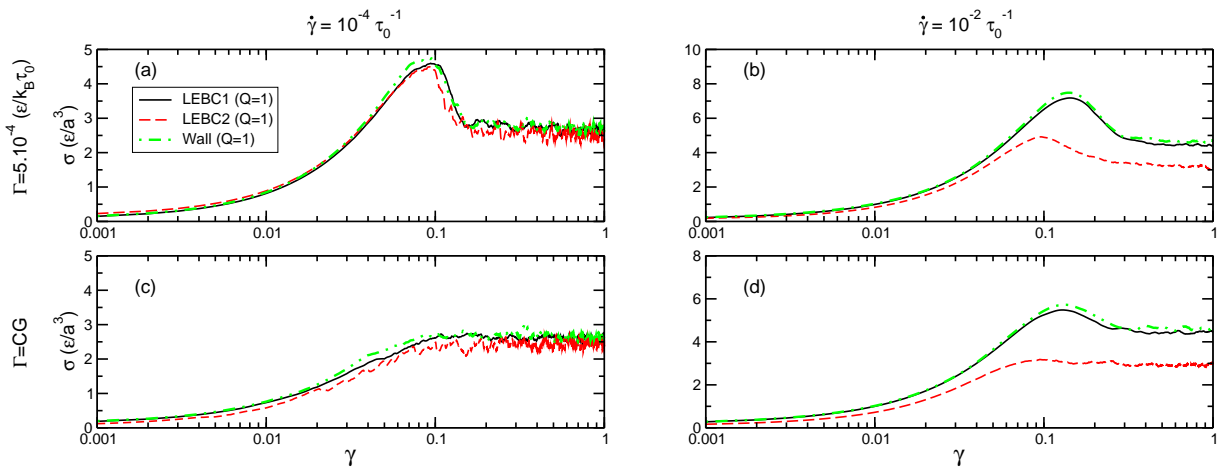}
\caption{Load curves obtained from uniform shearing at shear rate $\dot{\gamma}=1E-04 \tau_0^{-1}$ (left panel - (a) and (c)) and at $\dot{\gamma}=1E-02 \tau_0^{-1}$ (right panel - (b) and (d)) for initial configurations prepared at slow cooling rate (top panel) and at inifintely fast cooling rate (bottom panel). In each of the panel, the load curves are obtained from two different shear protocol, involving pair wise dissipation of drag (DPD protocol) and particle wise dissipation of drag (LD protocol). At low shear rates, both protocols show similar behaviour with the shear stress as a function of strain shows an initial elastic response, followed by an overshoot in stress, which eventually decays to reach a steady state value. The effect of dissipation is quite evident at the higher shear rate}.
\label{Fig5}
\end{figure*}

In Fig. \ref{Fig5} we show load curves, i.e., the shear stress $\sigma$ {\it vs.} applied strain $\gamma$, for the three different protocols (LEBC1, LEBC2 and WB) and for the two different samples (the well annealed one and the poorly annealed one mentioned above), sheared at $\dot{\gamma}=10^{-4} \invt$, a relatively low shear rate, and $\dot{\gamma}=10^{-2} \invt$, a higher shear rate. At low shear rate, all the protocols show comparable load curves. For the well annealed sample, independent of the protocol used, we find that the initial linear regime is followed by a stress overshoot before reaching a steady state value (see Fig. \ref{Fig5} (a)). The poorly annealed sample do not show any overshoot (see Fig. \ref{Fig5}). At relatively high shear rates (see Fig. \ref{Fig5} (b) and (d)), load curves consistently show the dependence on preparation protocol w.r.t the stress overshoot, but we also note that the dependence on shear protocol become more prominent. We also note that, approaching the steady state, the stress fluctuations are larger in the case of LEBC2 compared to LEBC1 and WB. When comparing LEBC1 and WB at high rate, we note that the difference in the boundary conditions does not seem to affect the value of the stress overshoot nor of the steady-state value, which instead change with changing the damping mechanism (LEBC1 vs LEBC2). 
\begin{figure}[t]
\includegraphics[width=0.40\textwidth]{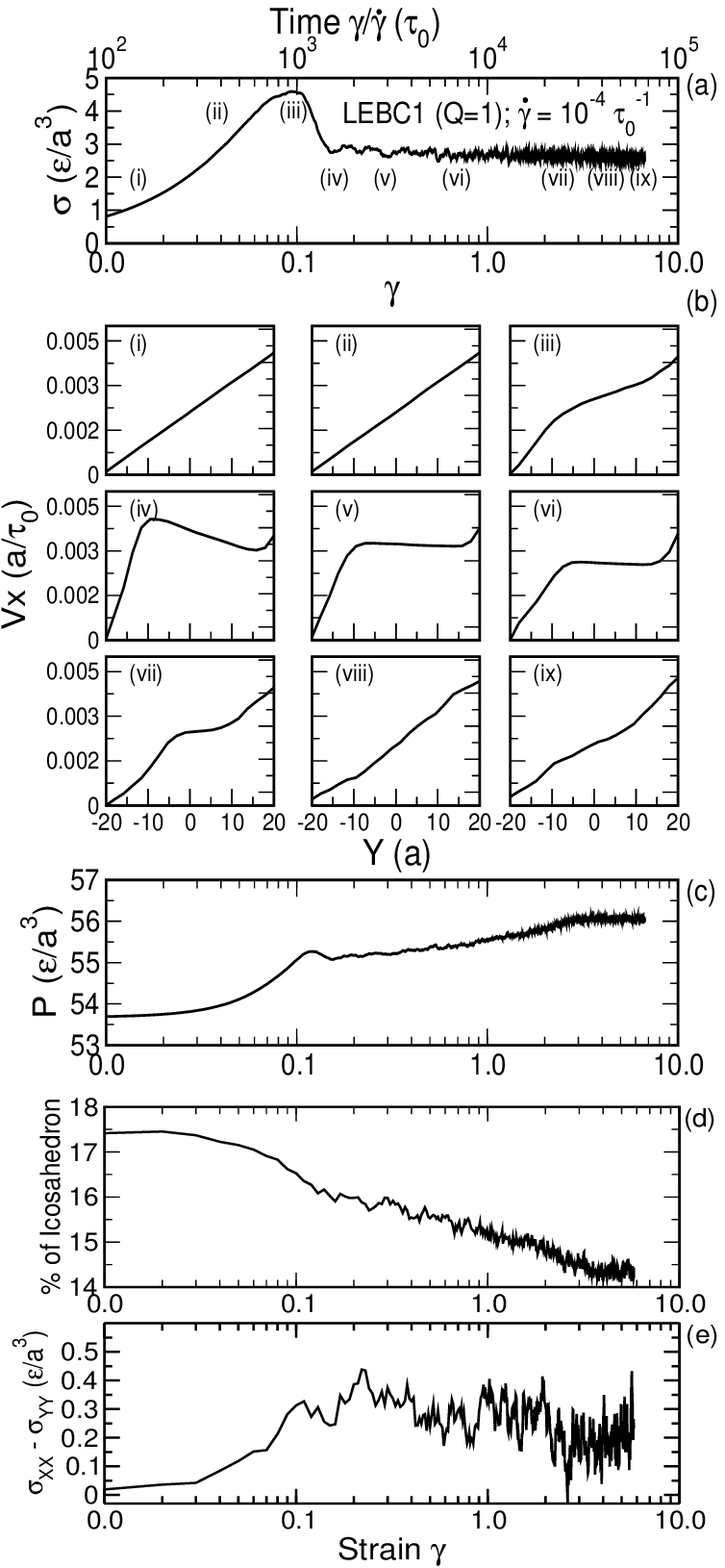}
\caption{(a) The load curve obtained from uniform shearing, using the LEBC1 (pair-wise dissipation) protocol, for Q=1, at shear rate $\dot{\gamma}=10^{-4} \invt$ for a sample initially prepared at cooling rate $\Gamma=5\cdot10^{-4} \epsilon/(k_{B} \tau_{0})$. (b) The velocity profile as a function of the coordinate in the gradient direction at points (i-ix) of the load curve. (c) The variation of pressure as a function of strain over the same range of strain as in the load curve. (d) The first normal stress difference ($\sigma_{\mathrm{XX}} - \sigma_{\mathrm{YY}}$) as a function of strain for the same sample. (e) The fraction of icosahedral packing as a function of strain.}
\label{Fig6}
\end{figure}
Now we focus our discussion on the low shear rate case ($\dot{\gamma}=10^{-4}\invt$) and on the well annealed samples, to analyze the corresponding velocity profiles. Figs. \ref{Fig6}, \ref{Fig7} and \ref{Fig8} show the velocity profiles respectively obtained with the LEBC1, LEBC2 and WB protocols from the same initial configurations. For different regimes in the load curves (Figs. \ref{Fig6} (a), \ref{Fig7}(a) and \ref{Fig8}(a)), we show the associated velocity profiles in Figs. \ref{Fig6}, \ref{Fig7} and \ref{Fig8} (b) (i-ix). The velocity profiles are computed over a strain window of $\Delta \gamma=0.02$. In spite of some differences in the details of the velocity profiles, we can recognize the following features for all the shearing protocols considered. 1) In the linear response regime of the load curve (i-ii), the velocity profile show a homogeneous flow with local shear rates (obtained as the slope of the velocity profile) similar to that of applied shear rate. 2) In the vicinity of the stress overshoot (iii) a flow instability develops, as indicated by the deviation from the linear velocity profile. Following the stress overshoot, the stress decays back, with $d\sigma/d\gamma < 0$ and in this region we find a back-flow in the system as the velocity profile show a negative slope \cite{moorcroft2013criteria, fielding_RPP_2014}. We note that the width of the shear band seems to be set by the amount of back flow in the system (iv). 3) Upon further shearing, the width of the band grows and eventually the whole system flow homogeneously. The shear component of the stress has a much weaker (although non-negligible) dependence on the strain and the pressure evolution with $\gamma$ clearly shows that saturation of the stress tensor only happens once the profile returns to be homogeneous (Figs. \ref{Fig6}, \ref{Fig7} and \ref{Fig8} (c)) \cite{olmsted_rheoacta_2008, divoux_prl_2010}. 

When analyzing the local packing through the Voronoi tessellation, we find that the fraction of icosahedrally packed particles has a similar trend. Fig. \ref{Fig6} (d) shows, for the LEBC1 protocol, how the fraction of icosahedrally packed particles  evolves with the increasing strain and the time required for it to saturate is consistent with the vanishing of the transient shear banding. The connections between the evolution of the local packing and that of the shear inhomogeneities, as well as their dependence on the shear rate have been thoroughly investigated in \cite{vasisht2017emergence}. We also find that the evolution of the system during the shear banding is accompanied by a positive first normal stress difference ($\sigma_{XX} - \sigma_{YY}$), indicating the presence of dilation, which, although quite noisy, also seems to grow as the bands develop and saturate as the flow becomes homogeneous (see Fig. \ref{Fig6} (e)). When comparing further the different protocols across Figs. \ref{Fig6} - \ref{Fig8}, we note that in WB the shear bands always nucleate near the walls and that the time taken to reach an homogeneous flow state is slightly longer than in LEBC1 and LEBC2. 

\begin{figure}[t]
\includegraphics[width=0.400\textwidth]{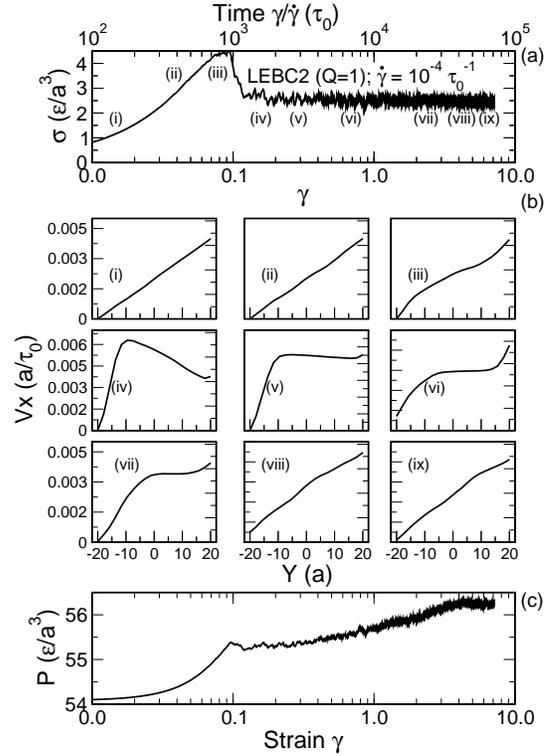}
\caption{(a) The load curve obtained from uniform shearing, using the LEBC2 protocol (Stoke's like dissipation), for Q=1, at shear rate $\dot{\gamma}=10^{-4} \invt$ for a initial configuration prepared at cooling rate $\Gamma=5\cdot10^{-4} \epsilon/(k_{B} \tau_{0})$. (b) The velocity profile as a function of the coordinate in the gradient direction at points (i-ix) of the load curve. (c) The variation of pressure as a function of strain over the same range of strain as in the load curve. Similar to the results of DPD protocol, the system shows the feature of transient shear banding as well the decoupling between the stress and pressure.}
\label{Fig7}
\end{figure}

We conclude that the features identified above, the stress overshoot, the band formation and disappearing, its correlation with the evolution of the normal stresses and with the changes in the local packing, are robust across the different protocols utilized. Hence they must not be the result of numerical artifacts and are instead inherent of well aged jammed solids sheared at sufficiently low rate.  
\begin{figure}[t]
\includegraphics[width=0.400\textwidth]{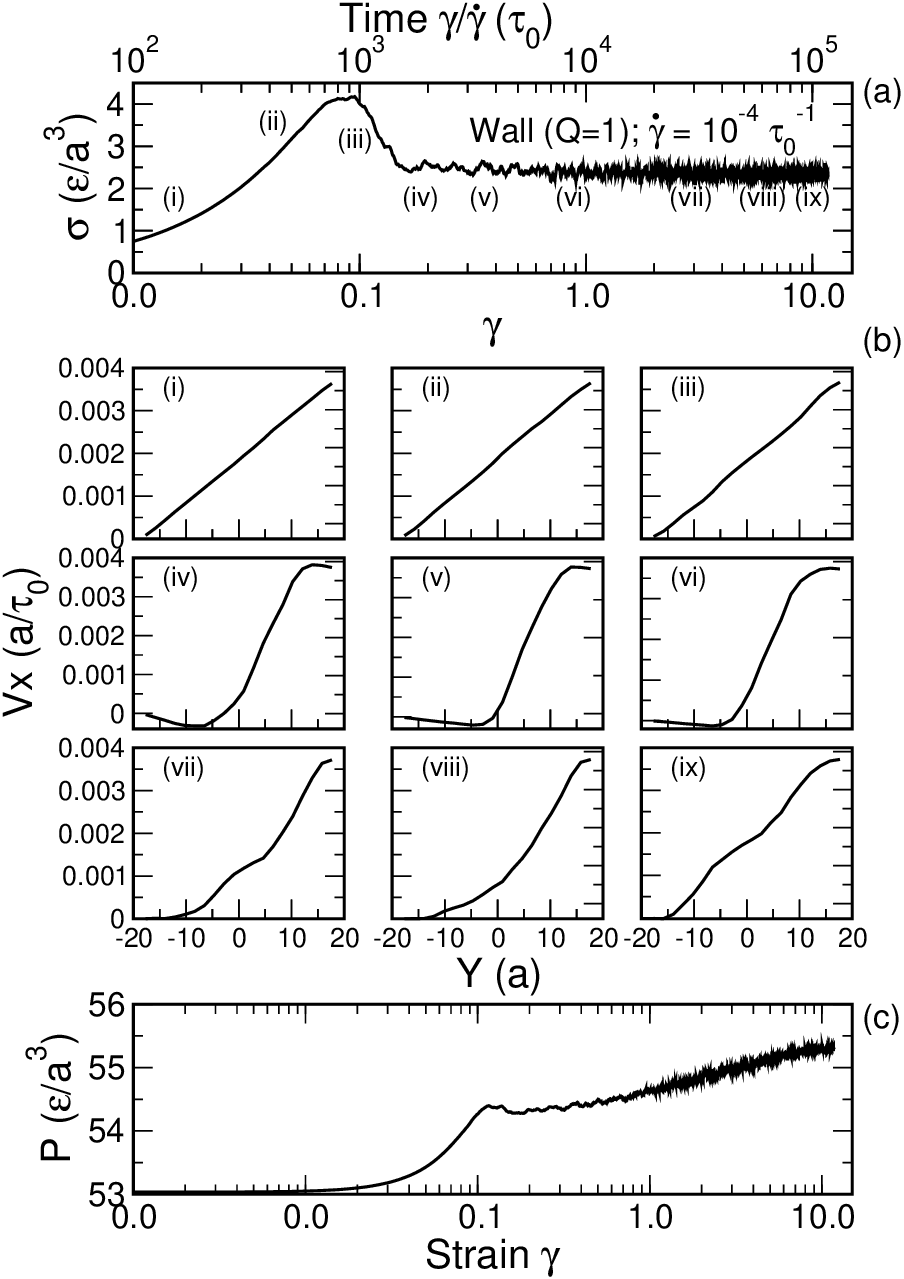}
\caption{(a) The load curve obtained from uniform shearing, using the wall based protocol, at shear rate	$\dot{\gamma}=10^{-4} \invt$ for a initial configuration prepared at cooling rate $\Gamma=5\cdot10^{-4} \epsilon/(k_{B} \tau_{0})$. (b) The velocity profile as a function of the coordinate in the gradient direction at points (i-ix) of the load curve. (c) The variation of pressure as a function of strain over the same range of strain as in the load curve.}
\label{Fig8}
\end{figure}

In the next section we discuss the possible dependence of our findings on the value of the drag coefficient chosen. 

\section{Dependence on dissipation co-efficient}
\begin{figure*}
\includegraphics[width=0.900\textwidth]{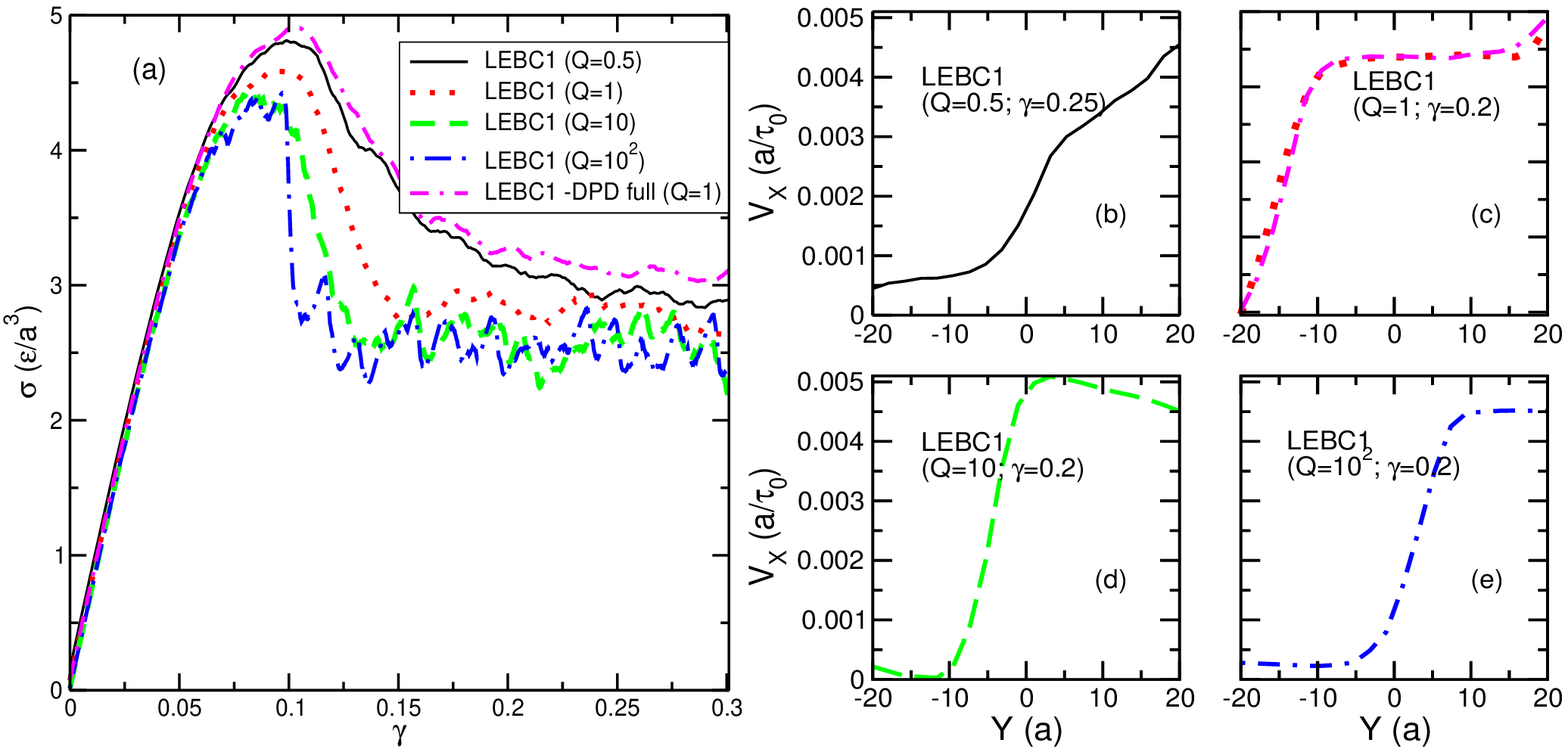}
\caption{(a) The load curve obtained from uniform shearing using different dissipation constants, at a shear rate $\dot{\gamma}=10^{-4} \invt$, for a initial configuration prepared at $\Gamma=5\cdot10^{-4} \epsilon/(k_{B} \tau_{0})$. Right panel (b-e) shows the velocity profile computed in a strain window $\gamma \approx 0.2$, averaged over $2\%$ strain for systems sheared using different dissipation constants. Even in the systems which are relatively underdamped, one finds the formation of shear bands at the vicinity of decay of stress overshoot. These bands are indeed transient in nature and the damping coefficient would have a bearing on the time required to obtain a homogeneous flow.}
\label{Fig9}
\end{figure*}

The contribution of inertial terms to the rheology of jammed suspensions has been extensively explored in steady-state in  \cite{salerno_PRL_2012, nicolas_PRL_2016, vasisht2018permanent}. Here we address instead how the stress overshoot and the shear banding during the transient preceding the steady state is affected by the dissipation co-efficient. To do this, we consider the LEBC1 protocol and vary the value of the coefficient $\zeta_{DPD}$ between $2$ ($m\invt$) and $0.01$ ($m\invt$). This means that the inertial quality factor $Q$ defined above varies from 0.5 to 100, i.e. from the case where the damping time is much shorter than the characteristic time of the inertial motion (i.e., the overdamped limit) to a case where it is ~100 times longer (underdamped case). In \cite{nicolas_PRL_2016, vasisht2018permanent} it has been shown that a strongly underdamped dynamics may qualitatively change the flow properties of the material, with the possibility to develop even non-monotonic flow curves. This is not the case here, where the flow curves remain monotonic, suggesting that for the whole range of $Q$ considered here the high density and the jammed conditions control the material flow, rather than the particle inertia. In Fig. \ref{Fig9} (a) we show the load curves of the same well annealed sample studied with LEBC1, but now with different values of the inertial quality factor Q. The applied shear rate corresponds to $10^{-4}\invt$. The data show that the initial linear response remains unaffected and the stress overshoot seems to slightly decrease with the increasing Q. The decay after the overshoot becomes increasingly steeper with increasing $Q$.  
If we express the applied shear rate in terms of Weissenberg number (${\mathrm W_i}=\zeta_{DPD} a^2/\epsilon \dot{\gamma}$), the higher the quality factor, the lower is $W_i$. As a consequence, one could think of the increasing quality factor $Q$ as a way to reach effectively lower shear rates. One might expect a trivial scaling of the load curves if the load curve is presented in terms of $\sigma$ {\it vs.} $\gamma / \dot{\gamma}$, with $\dot{\gamma}$ in terms of Weissenberg number. But since the initial linear response regime is unaffected by the extent of damping, this scaling would not work. The effect of the inertial contribution can be more intricate, with inertia playing a role in increasing the kinetic temperature in the athermal system and hence leading to a softening of the system during shearing \cite{nicolas_PRL_2016, vasisht2018permanent}.

In Figs. \ref{Fig9} (b)-(e), we show the velocity profiles related to different $Q$ and computed at a strain $\gamma\approx0.2$ by averaging over a strain window $\Delta \gamma=0.02$. In all the cases, as the stress decays from the overshoot, we observe the formation of shear bands. 

Finally, in the data of Fig. \ref{Fig9} (a) we also show the results obtained in the protocol LEBC1 when including the transverse contribution to the DPD drag, as discussed in section III, to demonstrate that such modification changes slightly the value of the overshoot and the shape of the decay towards the steady state, but it does not modify the general phenomenology of formation of shear band observed (see also the profile shown in Fig. \ref{Fig9} (c)). 

\begin{figure}
\includegraphics[width=0.400\textwidth]{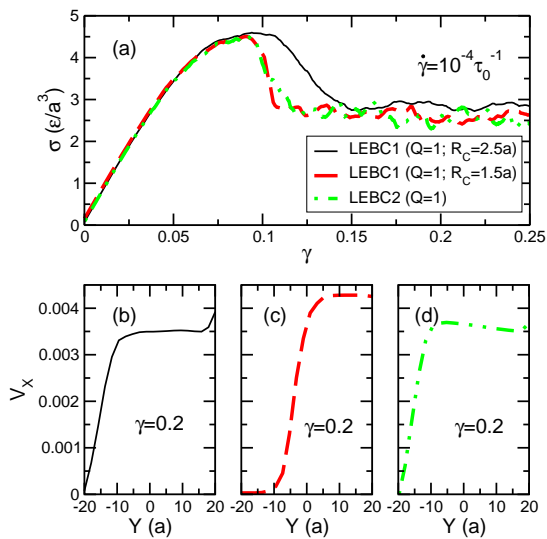}
\caption{(a) The load curve obtained from uniform shearing, using DPD cutoff $Rc=2.5a$ and $Rc=1.5a$, as well as using LD protocol, at a shear rate $\dot{\gamma}=10^-4 \invt$, for a initial configuration prepared at $\Gamma=5\cdot10^{-4} \epsilon/(k_{B} \tau_{0})$. Bottom panel (c-d) shows the velocity profile, corresponding to protocols shown in the load curve, computed in a strain	window $\gamma \approx 0.2$, averaged over $2\%$ strain.}
\label{Fig10}
\end{figure}

With the idea to explore further how the specific form of drag used may change the results obtained, we note that in LEBC1 another important variable corresponds to the pair-wise dissipation cutoff used in the DPD drag. For the results shown so far, we have chosen such cut-off to be $2.5\mathrm{a_{ij}}$, as done in \cite{nicolas_PRL_2016}. In Fig. \ref{Fig10} (a), we plot together the load curves obtained for LEBC1 with the DPD cut-off $2.5 a_{ij}$ and $1.5 a_{ij}$, along with the results for the LEBC2 protocol. We observe that the linear response regime is unaffected and so is the values of the stress overshoot. The stress decay from the the overshoot shows instead a dependence on the cut-off chosen. We observe that the results obtained with the LEBC2 protocol approach those for LEBC1 if the DPD cut-off distance goes down towards a particle diameter, consistent with the fact that LEBC2 corresponds to the free draining approximation. The related velocity profiles are shown in Figs. \ref{Fig10} (b)-(d) (these are computed at $\gamma=0.2$ and averaged over $\Delta \gamma=0.02$), indicating that the decay of the overshoot is always associated to the formation of transient shear bands for both values of the DPD cut-off.

\begin{figure}
\includegraphics[width=0.400\textwidth]{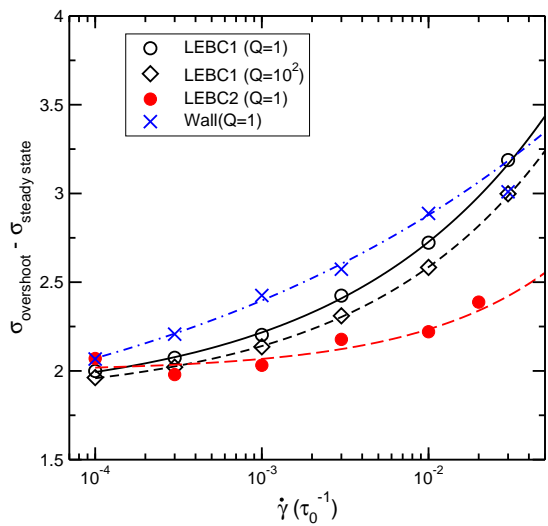}
\caption{The difference between the overshoot stress and steady state stress computed from different shearing protocols plotted as a function of shear rate. The lines represent the power law fit to the data points.}
\label{Fig11}
\end{figure}

Finally, we have also analyzed how the difference or similarities just described depend on the shear rate. We summarize the outcome of this study in Fig. \ref{Fig11}, which shows, for the well annealed sample, the difference between the stress overshoot and the steady state value of the shear stress as a function of the shear rate, for different protocols as well as for two different damping coefficients for LEBC1. The data indicate that changing $Q$ in the range of values explored here does not significantly affect the results at sufficiently low rates, whereas one should expect to see significant differences upon increasing the shear rate. Note that we have expressed the shear rate in terms of $\invt$, instead of using a viscous time scale, in order to compare different protocols where viscous terms are handled in different ways. Different protocols show that the occurrence of the stress overshoot (which is essential for the formation of shear bands) is robust and its dependence on the steady-state stress across different protocols is similar in the whole range of shear rates (the lines through the data in the figure are power law fits). Hence the stress overshoot and the banding observed at low rates are robust features of the systems we are studying, they are genuinely representative of the emerging physics and not the result of artifacts in the numerical simulations.

\section{Dependence on the sample age}
\begin{figure}
\includegraphics[width=0.400\textwidth]{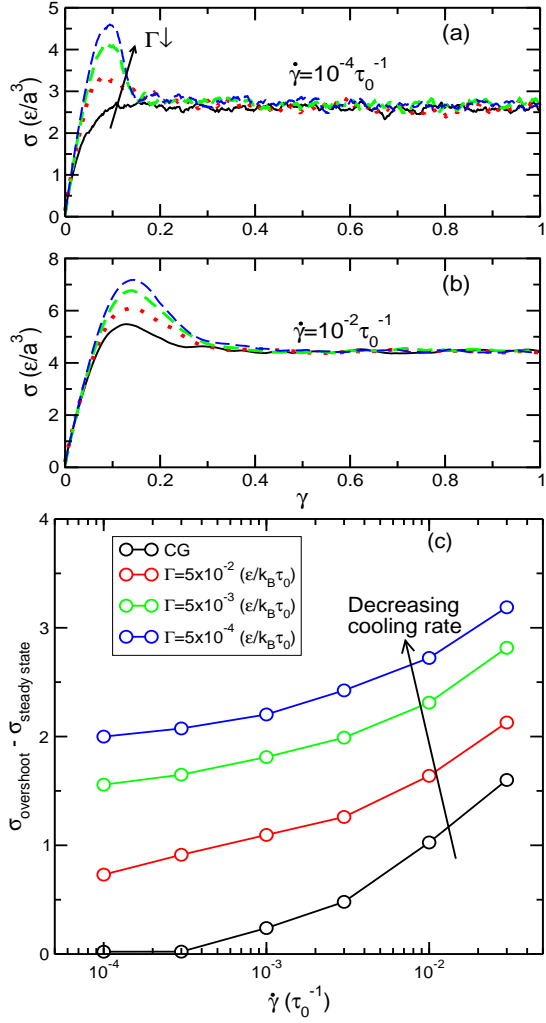}
\caption{The load curve for (a) $\dot{\gamma}=10^{-4}\invt$ and (b) $\dot{\gamma}=10^{-2}\invt$ computed from samples
prepared with different cooling rate $\Gamma$. (c) The difference in the overshoot stress and the steady state stress computed for different sample ages as a function of the applied shear rate.}
\label{Fig12}
\end{figure}

At the end this study, we want to emphasize that the tendency to have flow inhomogeneity upon yielding is determined mainly by the age of the samples (which determines the stress overshoot and eventual decay to steady state), due to the different degree of frozen-in stresses. In the context of the study reported here, we have analyzed the dependence of the stress overshoot on the shear rate and of the sample age. Fig. \ref{Fig12} shows the difference between the stress overshoot and the steady state value of the shear stress as a function of the shear rate, for different cooling rates $\Gamma$, corresponding to different sample ages. The results clearly show how changing the sample age can qualitatively change the overshoot and its rate dependence. We have investigated how the age of the samples also determine the persistence of the flow inhomogeneities, not only their presence, in the companion letter \cite{vasisht2017emergence}. We refer the reader to that paper for further insights.

\section{Conclusions}
We have devised a 3D numerical study of a jammed suspension of soft spheres, polydisperse in size, under shear. In particular here we have explored different choices for imposing the shear deformation and boundary conditions. We have compared the use of Lees-Edwards boundary conditions with simulations where the samples are confined within walls. We have also compared the use of a DPD drag term to the free draining approximation. Finally, we have compared simulations with different degrees of inertia, quantified through the inertia quality factor $Q$. The comparison has been done in terms of load curves and velocity profiles during the transient that leads to the steady-state flow. In all cases, we find that at low rates the shear stress develops an overshoot followed by a relatively long decay (not necessarily gradual) towards the steady state value. Such phenomenon is associated to a transient banding with a part of the material that is basically stuck, and the rest flowing. On the basis of the extended comparison carried on here, we propose that these features (the stress overshoot and the transient shear banding) are the genuine results of the emerging response of the material upon yielding and not the consequences of numerical artifacts or unphysical choices in the simulations parameters. We have been able to elucidate the dependence of the stress overshoot on the different choices for the simulation parameters and to clarify that what controls the response reported here is ultimately the age of the samples, which determines the spatial distribution of frozen-in stresses in the initial solid, as it has to be expected for jammed out-of-equilibrium materials. 

{\bf Acknowledgements.} The authors thank Georgetown University and the Swiss National Science Foundation (Grant No. PP00P2 150738) for support.

\clearpage
\bibliography{bibfile-Mar2019.bib}
\end{document}